\documentclass[conference]{IEEEtran}
\IEEEoverridecommandlockouts

\usepackage{cite}
\usepackage{amsmath,amssymb,amsfonts}
\usepackage{algorithmic}
\usepackage{graphicx}
\usepackage{textcomp}
\usepackage{xcolor}
\usepackage{multirow}
\usepackage{hyperref}
\usepackage[top=0.75in, bottom=1in, left=0.625in, right=0.625in]{geometry}

\def\BibTeX{{\rm B\kern-.05em{\sc i\kern-.025em b}\kern-.08em
    T\kern-.1667em\lower.7ex\hbox{E}\kern-.125emX}}
\begin{document}

\title{Towards Lightweight Hyperspectral Image Super-Resolution with Depthwise Separable Dilated Convolutional Network\\
}

\author{
    \IEEEauthorblockN{
        Usman Muhammad\textsuperscript{1}, Jorma Laaksonen\textsuperscript{1}, and Lyudmila Mihaylova\textsuperscript{2}
    }
    \IEEEauthorblockA{\textsuperscript{1} Department of Computer Science, Aalto University, Finland}
    \IEEEauthorblockA{\textsuperscript{2} School of Electrical and Electronic Engineering, University of Sheffield, United Kingdom}

}

\maketitle

\begin{abstract}
Deep neural networks have demonstrated highly competitive performance in super-resolution (SR) for natural images by learning mappings from low-resolution (LR) to high-resolution (HR) images. However, hyperspectral super-resolution remains an ill-posed problem due to the high spectral dimensionality of the data and the scarcity of available training samples. Moreover, existing methods often rely on large models with a high number of parameters or require the fusion with panchromatic or RGB images, both of which are often impractical in real-world scenarios. Inspired by the MobileNet architecture, we introduce a lightweight depthwise separable dilated convolutional network (DSDCN) to address the aforementioned challenges. Specifically, our model leverages multiple depthwise separable convolutions, similar to the MobileNet architecture, and further incorporates a dilated convolution fusion block to make the model more flexible for the extraction of both spatial and spectral features. In addition, we propose a custom loss function that combines mean squared error (MSE), an L2 norm regularization-based constraint, and a spectral angle-based loss, ensuring the preservation of both spectral and spatial details. The proposed model achieves very competitive performance on two publicly available hyperspectral datasets, making it well-suited for hyperspectral image super-resolution tasks. The source codes are publicly available at: \href{https://github.com/Usman1021/lightweight}{https://github.com/Usman1021/lightweight}.
\end{abstract}

\begin{figure*}[htbp]
    \centering
    \includegraphics[width=0.91\textwidth]{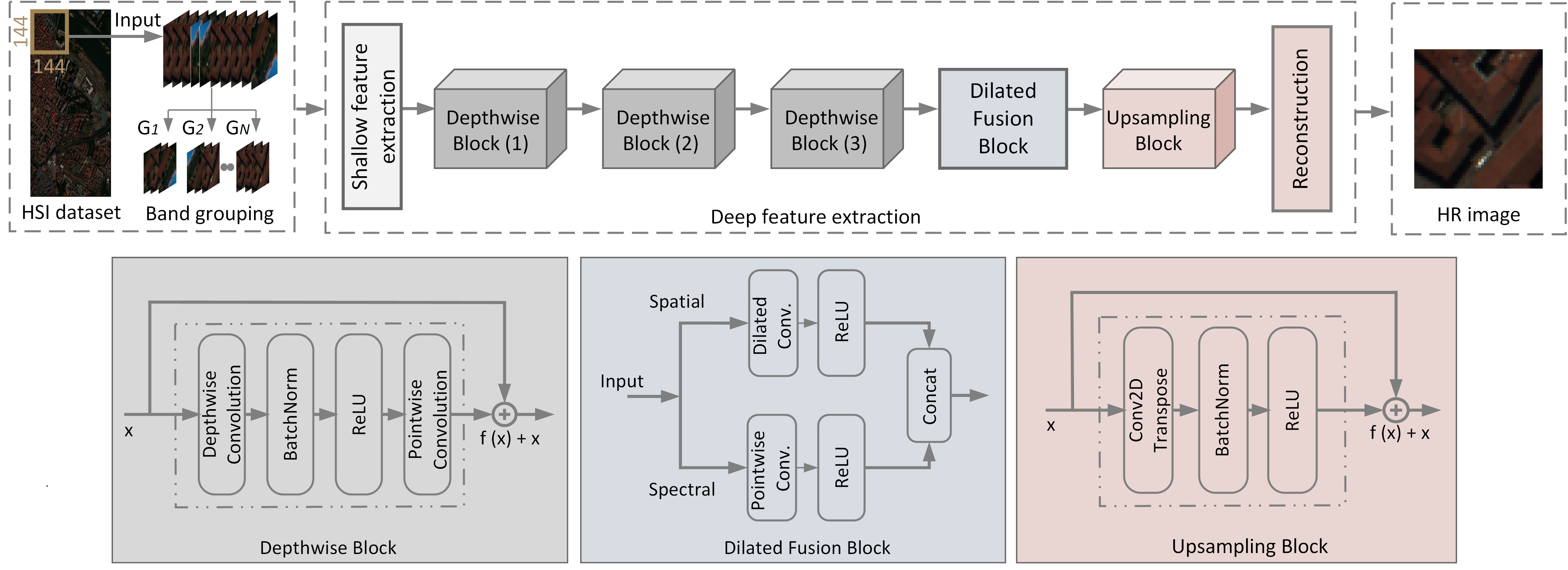} 
    \caption{A detailed overview of the proposed model. Shallow features are initially extracted using a lightweight separable convolutional layer with ReLU activation. The gray color represent depthwise convolutional blocks for deep feature extraction. The blue block denotes the dilated fusion module, capturing spatial and spectral features in parallel. The pink block performs upsampling to reconstruct the high-resolution output.}
    \label{fig}
\end{figure*}
\section{Introduction}
 Hyperspectral imaging systems capture surface information across numerous spectral bands, providing richer spectral details than multispectral or natural images. This enables precise characterization of ground objects \cite{singh2020hyperspectral,7284770,6151825,6151826}. Moreover, the wide spectrum of information makes them extremely valuable for a variety of applications, including anomaly detection, surveillance, environmental monitoring, and satellite image scene classification \cite{muhammad2018pre, muhammad2018feature, muhammad2019bag, muhammad2022patch, peng2022lcrca}. However, hyperspectral images often suffer from low resolution due to environmental factors and sensor energy limitations \cite{bu2024sbhsr}. Deep learning-based super-resolution (SR) methods have shown remarkable success, leveraging large models to reconstruct high-resolution images from degraded low-resolution inputs \cite{hou2022deep, chen2023msdformer}. Nonetheless, due to advancements in deep learning models, the number of parameters in deep learning networks has increased significantly. As a result, numerous lightweight super-resolution models have gained significant attention in recent years \cite{guo2024towards}, focusing on reducing model parameters and computational complexity through various strategies. 
 
In particular, knowledge distillation \cite{lee2020learning} is widely used for model compression, in which a large model (the teacher network) is first trained on the original data and then used to guide the intermediate feature representations of a smaller model (the student network). For instance, Gao et al. \cite{gao2018image} employ a lightweight student SR model to acquire knowledge from a deeper teacher SR network. Other model compression techniques, such as pruning \cite{zhan2021achieving} and low-rank factorization \cite{wu2018convex}, have also proven effective for developing super-resolution models. For instance, Zhan et al. \cite{zhan2021achieving} integrate neural architecture search with a layer-wise pruning strategy to develop efficient super-resolution model. Wu et al. \cite{wu2018convex} propose a convex regularizer for low-rank solutions, addressing unconstrained cases with proximal gradient and a custom PG-like algorithm. In addition, some quantization-based methods \cite{xin2020binarized, ma2019efficient } leverage low-bit representations to enhance the inference speed of super-resolution models. However, these models often exhibit a significant performance gap compared to state-of-the-art super-resolution methods \cite{zhang2024hyperspectral}. 

In order to reduce the size of the neural network, convolutional filters play a crucial role, and significant efforts have been made to maximize their potential. For instance, MobileNet \cite{howard2017mobilenets} model introduces depthwise separable convolution, demonstrating great potential for network decomposition. Ahn et al. \cite{ahn2018fast} employ group convolution instead of depthwise convolution to make model efficiency more adaptable. Kim et al. \cite{kim2016deeply} propose a deeply recursive convolutional network, arguing that increasing recursion depth can enhance performance without introducing additional parameters for extra convolutions. Dilated convolution is another modified version of standard convolutions that utilizes a distinctive grid structure \cite{hu2024hdconv}. It has been widely adopted to expand the receptive field of kernels and enhance global information aggregation, leading to notable improvements over previous methods. However, the classification accuracy of these models may be compromised during compression, as a large number of image features are omitted due to the simplified convolution operations \cite{sun2020lightweight}.

Our approach is inspired by MobileNet \cite{howard2017mobilenets}, which reduces the number of parameters and computational costs by introducing the concept of depthwise separable convolution. In particular, we aim to combine depthwise separable convolutions with dilated convolutional fusion blocks \cite{yu2015multi} to develop a novel super-resolution model called the depthwise separable dilated convolution network (DSDCN) for hyperspectral images. Specifically, the dilated convolutional fusion block consists of two different types of convolutions: (1) dilated convolution (spatial branch) and (2) pointwise convolution (spectral branch). The outputs of the dilated convolution and pointwise convolution are then concatenated to effectively integrate spatial and spectral information. By doing so, we make our model suitable for hyperspectral super-resolution tasks while minimizing the compromise on model parameters compared to the MobileNet architecture. To further improve the performance of the proposed method, we employ a custom loss function that integrates mean squared error (MSE), an L2 regularization-based constraint, and a spectral angle-based loss to ensure high-fidelity reconstruction. In summary, our contributions are three-fold:

\begin{enumerate}
    \item We present a novel lightweight DSDCN that combines depthwise separable convolutions, residual connections, and dilated convolutional fusion to enhance spatial resolution and preserve spectral integrity..
    \item A custom loss function is introduced, integrating mean squared error (MSE), an L2 regularization constraint, and a spectral angle-based loss to enhance high-fidelity reconstruction.
    \item Experiments on two hyperspectral datasets are conducted across various resolution degradation-restoration scenarios (2$\times$, 4$\times$, and 8$\times$ downsampling), demonstrating competitive performance on both datasets.
\end{enumerate}

\section{Methodology}
Fig. 1 shows the band grouping and the three main components of the model: (1) depthwise separable convolutions with residual connections, (2) dilated convolution fusion, and (3) upsampling. We first define the band grouping along with other components. The following subsections detail this and the custom loss used in the model.

\subsection{Band Grouping}
Since hyperspectral images consist of hundreds of spectral bands, processing all bands together can be computationally challenging and may introduce redundancy due to the high spectral correlation between bands. To mitigate this, we utilize band grouping \cite{wang2024enhancing}, which involves partitioning adjacent bands into overlapping groups for seamless integration with our proposed model. Specifically, hyperspectral bands are organized into overlapping subgroups by defining a fixed group size with a designated overlap, ensuring that consecutive subgroups share common bands.

\subsection{Depthwise Separable Convolution}
Given a low-resolution hyperspectral image \( X \in \mathbb{R}^{H \times W \times B} \), where \( H \), \( W \), and \( B \) denote the spatial height, width, and number of spectral bands, respectively, our goal is to reconstruct a high-resolution image \( \hat{X} \in \mathbb{R}^{\alpha H \times \alpha W \times B} \) with an upscaling factor \( \alpha \). To achieve this, we design a deep neural network \( \mathcal{F}(X; \theta) \) that learns the LR-to-HR mapping efficiently while preserving spectral integrity.

We begin by extracting spatial and spectral features using depthwise separable convolutions. Specifically, we empirically employ three such blocks, where each block decomposes a standard convolution into two sequential operations \cite{howard2017mobilenets}:
\begin{equation}
    X_d' = W_d * X,
\end{equation}
where \( W_d \) denotes the depthwise convolution kernel applied independently to each spectral band. Next, a pointwise convolution is applied to combine features across channels:
\begin{equation}
    X_p' = W_p * X_d',
\end{equation}
where \( W_p \) is a \( 1 \times 1 \) convolution kernel that processes each spatial location across all channels. To improve training stability and mitigate vanishing gradients, we incorporate a residual connection by applying a separate \( 1 \times 1 \) convolution to the original input \( X \) \cite{sandler2018mobilenetv2}:

\begin{equation}
    X_{\text{res}} = W_s * X,
\end{equation}
where \( W_s \) is the projection kernel for aligning dimensions. The final output of the block is obtained by fusing the residual and transformed features:
\begin{equation}
    C_{\text{out}} = X_p' + X_{\text{res}},
\end{equation}
where \( C_{\text{out}} \) is the output feature map of the depthwise separable convolution block.

\subsection{Dilated Convolution Fusion}

To integrate both local and global spatial dependencies while maintaining a lightweight structure, we employ a \textit{dilated fusion block} composed of three parallel \(3 \times 3\) convolutions with increasing dilation rates \( r = \{1, 2, 3\} \). Given an input feature map \( \mathbf{F}_{\text{in}} \in \mathbb{R}^{H \times W \times C_{\text{in}}} \), feature extraction is performed as \cite{yu2015multi}:
\begin{equation}
F_i = \text{ReLU}(Q_i * F_{\text{in}}), \quad i \in \{1, 2, 3\},
\end{equation}
where each \( Q_i \in \mathbb{R}^{3 \times 3 \times C_{\text{in}} \times C} \) is a convolutional kernel with dilation rate \( r_i \in \{1, 2, 3\} \), and \( C \) is the number of output channels per branch. The feature maps \( F_1 \), \( F_2 \), and \( F_3 \) are obtained by applying ReLU-activated convolutions with kernels \( Q_1 \), \( Q_2 \), and \( Q_3 \), respectively, to the same input \( F_{\text{in}} \), and are then concatenated along the channel dimension.
\begin{equation}
F_{\text{concat}} = \text{Concat}(F_1, F_2, F_3),
\end{equation}
The concatenated features are then passed through a final \(1 \times 1\) convolution to fuse the multi-scale information:
\begin{equation}
F_{\text{out}} = \text{ReLU}(Q_f * F_{\text{concat}}),
\end{equation}
where \( Q_f \in \mathbb{R}^{1 \times 1 \times 3C \times C} \) is the fusion kernel. The output \( F_{\text{out}} \in \mathbb{R}^{H \times W \times C} \) represents the final feature map enriched with contextual spatial features.

\subsection{Upsampling with Transpose Convolution}
To increase spatial resolution, we employ an upsampling block based on transpose convolution \cite{dong2016accelerating}. Given a low-resolution feature map \( U_{\text{in}} \in \mathbb{R}^{H \times W \times C_u} \), a learnable transpose convolution is defined as:
\begin{equation}
U_{\text{tr}} = K_{\text{tr}} * U_{\text{in}},
\end{equation}
where \( K_{\text{tr}} \in \mathbb{R}^{k \times k \times C_u \times C_o} \) is the transpose convolution kernel, and \( U_{\text{tr}} \in \mathbb{R}^{\alpha H \times \alpha W \times C_o} \) is the upsampled output. Here, \( C_u \) is the number of input channels, and \( C_o \) is the number of output channels after upsampling. The upsampled output is then passed through a non-linearity:
\begin{equation}
U_{\text{act}} = \text{ReLU}(U_{\text{tr}}).
\end{equation}
Furthermore, we employ a skip connection by projecting and upsampling the input feature map:
\begin{equation}
U_{\text{skip}} = K_{\text{skip}} * \text{Up}(U_{\text{in}}),
\end{equation}
where \( K_{\text{skip}} \in \mathbb{R}^{1 \times 1 \times C_u \times C_o} \) is a projection kernel. The final output is obtained by summing the skip and upsampled features:
\begin{equation}
U_{\text{out}} = U_{\text{act}} + U_{\text{skip}},
\end{equation}
where \( U_{\text{out}} \in \mathbb{R}^{\alpha H \times \alpha W \times C_o} \) is the resulting feature map.

\begin{table}[t]
\caption{Quantitative results and model complexity}
\label{tab:switched_model_comparison}
\centering
\resizebox{0.8\columnwidth}{!}{
\begin{tabular}{|c|c|c|}
\hline
\multicolumn{3}{|c|}{Ablation study on PaviaC (2×)} \\
\hline
Model Variant & MSSIM$\uparrow$ & MPSNR$\uparrow$ \\
\hline
DSDCN w/o band grouping & 0.9482 & 35.017 \\
\hline
DSDCN w/s 16 & 0.9558 & 35.447 \\
\hline
DSDCN + w/s 32 & \textbf{0.9578} & \textbf{36.434} \\
\hline
DSDCN w/s 48 & 0.9502 & 35.861 \\
\hline
DSDCN + w/o custom loss & 0.9507 & 36.127 \\
\hline
\multicolumn{3}{|c|}{} \\  
\hline
\multicolumn{3}{|c|}{Model Complexity} \\
\hline
Model & Scale & Parameters \\
\hline
ERCSR \cite{li2021exploring} & 4 & 1.59M \\
\hline
MCNet \cite{li2020mixed} & 4 & 2.17M \\
\hline
PDENet \cite{hou2022deep} & 4 & 2.30M \\
\hline
CSSFENet \cite{zhang2024hyperspectral} & 4 & 1.61M \\
\hline
DSDCN (Ours) & 4 & \textbf{0.96M} \\
\hline
\end{tabular}
}
\end{table}

\begin{table*}[tb]
\centering
\caption{Evaluation on datasets (PaviaC, PaviaU) in different scaling setups. The comparison results are reported from \cite{zhang2024hyperspectral}.}
\label{tab:results_paviac_paviau_centered_algorithm_scalefactor}
\scalebox{1.0}{
\begin{tabular}{|c|c|ccc|ccc|}
\hline
\multirow{2}{*}{\textbf{Scale Factor}} & \multirow{2}{*}{\textbf{Model}} & \multicolumn{3}{c|}{\textbf{PaviaC}} & \multicolumn{3}{c|}{\textbf{PaviaU}} \\ 
\cline{3-8}
                      &                   & \textbf{MPSNR$\uparrow$} & \textbf{MSSIM$\uparrow$} & \textbf{SAM$\downarrow$} & \textbf{MPSNR$\uparrow$} & \textbf{MSSIM$\uparrow$} & \textbf{SAM$\downarrow$} \\ \hline
\multirow{8}{*}{\centering $\boldsymbol{2\times}$} 
    & VDSR   \cite{kim2016accurate}     & 34.879 & 0.9501 & 3.689  & 34.038  & 0.9524 & 3.258 \\ 
    & MCNet   \cite{li2020mixed}    & 34.626 & 0.9455 & 3.865  & 33.743  & 0.9502 & 3.359 \\ 
    & EDSR   \cite{lim2017enhanced}    & 34.580 & 0.9452 & 3.898  & 33.985  & 0.9511 & 3.334 \\ 
    & MSDformer    \cite{chen2023msdformer}    & 35.028 & 0.9493 & 3.691  & 34.159  & 0.9553 & 3.211 \\ 
    & MSFMNet  \cite{zhang2021multi}           & 35.200 & 0.9506 & 3.656  & 34.980  & 0.9582 & 3.160 \\ 
    & AS3 ITransUNet    \cite{xu20233}   & 35.221 & 0.9511 & 3.612  & 35.163  & 0.9591 & 3.149\\ 
    & PDENet    \cite{hou2022deep}      & 35.244 & 0.9519 & 3.595  & 35.275  & 0.9594 & 3.142 \\ 
    & CSSFENet  \cite{zhang2024hyperspectral}     & 35.522 & 0.9544 & 3.542  & 35.924  & \textbf{0.9625}  & \textbf{3.038} \\ 
    & DSDCN (Ours) & \textbf{36.434} &  \textbf{0.9578} & \textbf{3.538} & \textbf{35.941} & 0.9442 & 3.703 \\ \hline
\multirow{8}{*}{\centering $\boldsymbol{4\times}$} 
    & EDSR  \cite{lim2017enhanced}    & 28.591&0.7782&6.573  & 29.894 &0.7791 &5.074 \\ 
    & VDSR \cite{kim2016accurate}      & 28.317&0.7707&6.514  & 29.904&0.7753&4.997 \\ 
    & MCNet   \cite{li2020mixed}   & 28.756&0.7826&6.385  & 29.993&0.7835&4.917 \\ 
    & MSDformer   \cite{chen2023msdformer}      & 28.810&0.7833&5.897  & 30.098&0.7905&4.885 \\ 
    & MSFMNet  \cite{zhang2021multi}        &   28.873& 0.7863&6.300  & 30.283&0.7948&4.861 \\ 
    & AS3 ITransUNet  \cite{xu20233}  &  28.874&0.7893&5.972  & 30.289& 0.7940&4.859\\ 
    & PDENet  \cite{hou2022deep}       & 28.951&0.7900&5.876  & 30.295& 0.7944& 4.853 \\ 
    & CSSFENet   \cite{zhang2024hyperspectral}   & 29.054 & 0.7961 & 5.816  & \textbf{30.689} & \textbf{0.8107} & 4.839 \\ 
    & DSDCN  (Ours)       & \textbf{29.665} & \textbf{0.8152} & \textbf{4.826} & 30.524 & 0.7958 & \textbf{4.807} \\ \hline
\multirow{8}{*}{\centering $\boldsymbol{8\times}$} 
    & VDSR  \cite{kim2016accurate}     & 24.804&0.4944&7.588  & 27.028&0.5962&7.133 \\ 
    & EDSR   \cite{lim2017enhanced}   & 25.067&0.5282&7.507 & 27.467&0.6302&6.678 \\ 
    & MCNet  \cite{li2020mixed}    & 25.096&0.5391&7.429  & 27.483&0.6254&6.683 \\ 
    & MSDformer  \cite{chen2023msdformer}       & 25.215&0.5462&7.427  & 27.323&0.6341&6.668 \\ 
    & MSFMNet  \cite{zhang2021multi}        &   25.257&0.5464&7.449  & 27.586&0.6356&6.615 \\ 
    & AS3 ITransUNet  \cite{xu20233}  &  25.258&0.5435&7.417  & 27.689&0.6413&6.574\\ 
    & PDENet   \cite{hou2022deep}      & 25.288&0.5436&7.402  & 27.738&0.6457&6.531 \\ 
    & CSSFENet   \cite{zhang2024hyperspectral}   & 25.359 & 0.5493 & 7.306   & 27.825 &  \textbf{0.6569} & 6.505 \\ 
    & DSDCN  (Ours)        & \textbf{25.463} & \textbf{0.5553} & \textbf{6.345} & \textbf{27.827} & 0.6235 & \textbf{6.487} \\ \hline
\end{tabular}
}
\end{table*}
\subsection{Custom Loss Function}
We employ a custom loss function that combines three components: mean squared error (MSE), spectral angle mapper (SAM) loss, and \( \ell_2 \) loss. This combination ensures both pixel-wise accuracy and spectral fidelity in hyperspectral reconstruction. The total loss is defined as:
\begin{equation}
\mathcal{L}_{\text{total}} = \mathcal{L}_{\text{MSE}} + \lambda_1 \cdot \mathcal{L}_{\text{SAM}} + \lambda_2 \cdot \mathcal{L}_{\ell_2},
\end{equation}
where: \( \mathcal{L}_{\text{MSE}} = \frac{1}{N} \sum_{i=1}^{N} (y^{(i)}_{\text{true}} - y^{(i)}_{\text{pred}})^2 \) minimizes pixel-wise differences, \( \mathcal{L}_{\text{SAM}} = \frac{1}{N} \sum_{i=1}^{N} \cos^{-1}\left( \frac{\langle y^{(i)}_{\text{true}}, y^{(i)}_{\text{pred}} \rangle}{\|y^{(i)}_{\text{true}}\| \cdot \|y^{(i)}_{\text{pred}}\|} \right) \) measures the angular discrepancy between predicted and ground-truth spectral vectors, \( \mathcal{L}_{\ell_2} = \frac{1}{N} \sum_{i=1}^{N} \| y^{(i)}_{\text{true}} - y^{(i)}_{\text{pred}} \|_2^2 \) reinforces global spectral consistency.
Here, \( \lambda_1 \) and \( \lambda_2 \) are weighting coefficients that balance the influence of SAM and \( \ell_2 \) loss, respectively. In our experiments, we set \( \lambda_1 = 0.5 \) and \( \lambda_2 = 0.03 \).

\section{Experimental setup}

\subsection{Datasets and Implementation}
Two publicly available hyperspectral datasets, PaviaC and PaviaU, containing 102 and 103 spectral bands, respectively, are used in our study. We used a patch size of $144 \times 144$, following the protocol of previous work \cite{zhang2024hyperspectral}, for training and testing the datasets. Specifically, the test set patch is extracted from the bottom center of the PaviaC dataset and the top left of the PaviaU dataset, while the remaining image regions are used for training \cite{zhang2024hyperspectral}. 

To generate low-resolution images, we employ area-based sampling and downscale them by factors of 2×, 4×, and 8×. During training, the Adam optimizer was employed with a batch size of 4. Additionally, an early stopping function was used to prevent fixed epochs and avoid overfitting. Both datasets were divided into band groups of size 32, with one-fourth overlap between them. We adopted several widely used metrics to evaluate the quality of reconstructed images quantitatively, including mean peak signal-to-noise ratio (MPSNR),  mean structural similarity index (MSSIM), and Spectral Angle Mapper (SAM)\cite{chudasama2024comparison}.

\subsection{Ablation Study on Band Grouping}
Table I presents an ablation study evaluating the effects of band grouping size and the custom loss function on the performance of the proposed DSDCN model. The baseline without band grouping yields an MSSIM of 0.9482 and an MPSNR of 35.017 dB. It can be observed that using a band grouping size of 16 improves performance, while a grouping size of 32 achieves the best results (MSSIM: 0.9578, MPSNR: 36.434 dB). However, increasing the group size to 48 slightly reduces accuracy, suggesting that excessive grouping may weaken the spatial-spectral feature alignment. Performance is also reported using only the MSE loss (without the custom loss), with the group size fixed at 32. This results in a performance drop (MSSIM: 0.9507, MPSNR: 36.127 dB), highlighting the benefits of incorporating spectral angle and L2 constraints into the loss function.

\subsection{Comparison with State-of-the-Art Methods}
As shown in Table I, we compare the parameter count of our proposed model with several state-of-the-art super-resolution methods, including ERCSR (1.59M) \cite{li2021exploring}, MCNet (2.17M) \cite{li2020mixed}, PDENet (2.30M) \cite{hou2022deep}, and CSSFENet (1.61M) \cite{zhang2024hyperspectral}. DSDCN contains only 0.96 million parameters, substantially fewer than those of the other methods. 

Table II presents a detailed quantitative comparison under multiple scaling factors (2×, 4×, and 8×) on both the PaviaC and PaviaU datasets. The comparison includes various state-of-the-art methods such as VDSR \cite{kim2016accurate}, EDSR \cite{lim2017enhanced}, MCNet \cite{li2020mixed}, MSDformer \cite{chen2023msdformer}, MSFMNet \cite{zhang2021multi}, AS3 ITransUNet \cite{xu20233}, PDENet \cite{hou2022deep}, and CSSFENet \cite{zhang2024hyperspectral}. Despite its lightweight nature, our model ranks first in the PaviaC dataset and achieves competitive performance on the PaviaU dataset in terms of PSNR and SAM. These results emphasize the strength of our architecture in capturing both spatial and spectral information for hyperspectral image super-resolution.

\section{Conclusion}
In this study, we introduced a lightweight depthwise separable dilated convolutional network (DSDCN) for hyperspectral super-resolution. The model leverages depthwise separable convolutional blocks with a dilated convolution fusion mechanism to improve spatial and spectral feature extraction. Additionally, we proposed a custom loss function that integrates MSE, L2 regularization, and a spectral angle-based loss to ensure high-fidelity reconstruction. Experimental results demonstrate that our approach achieves competitive performance on two publicly available hyperspectral datasets while maintaining a relatively compact model size. This makes our method a practical and efficient solution for real-world hyperspectral image super-resolution tasks.

\section*{Acknowledgment}
This project has been funded by the European Union’s NextGenerationEU instrument and the Research Council of Finland under grant \textnumero{} 348153, as part of the project \emph{Artificial Intelligence for Twinning the Diversity, Productivity and Spectral Signature of Forests} (ARTISDIG). We also gratefully acknowledge CSC for providing access to the LUMI supercomputer, operated by the EuroHPC Joint Undertaking.

\vspace{12pt}

\bibliographystyle{IEEEtran}
\bibliography{references}

\end{document}